\documentclass{aa}
\usepackage{graphics}

\def\ltsima{$\; \buildrel < \over \sim \;$}
\def\lsim{\lower.5ex\hbox{\ltsima}}
\def\gtsima{$\; \buildrel > \over \sim \;$}
\def\gsim{\lower.5ex\hbox{\gtsima}}

\begin{document}

\thesaurus{3 
           (02.01.2; 
            11.01.2; 
            11.14.1; 
            11.17.3; 
            12.07.1; 
            )}   

\title{Microlens diagnostics of accretion disks \protect\\
in active galactic nuclei}

\author{A. Yonehara\inst{1}\inst{2} \and S. Mineshige\inst{1} 
\and J. Fukue\inst{3} \and M. Umemura\inst{4} \and E.L. Turner\inst{5}}

\offprints{A. Yonehara}
\mail{yonehara@kusastro.kyoto-u.ac.jp}

\institute{Department of Astronomy, Kyoto University, Sakyo-ku, 
Kyoto 606-8502, Japan 
 \and 
Research Fellow of the Japan Society for the Promotion of Science
 \and
Astronomical Institute, Osaka Kyoiku University,
Asahigaoka, Kashiwara, Osaka 582-0026, Japan
 \and 
Center for Computational Physics, Tsukuba University,
Tsukuba, Ibaraki 305-0006, Japan
 \and 
Princeton University Observatory, Peyton Hall, Princeton, NJ 08544, USA }

\date{Received ~~~~~~~~ / Accepted ~~~~~~~~ }

\maketitle

\begin{abstract}
The optical-ultraviolet continuum from active galactic nuclei (AGN)
seems to originate from optically thick and/or thin disks, and
occasionally from associated circumnuclear starburst regions. 
These different possible origins can, in principle, be discriminated 
by observations of gravitational microlensing events.
We performed numerical simulations of microlensing of an AGN
disk by a single lensing star passing in front of the AGN. 
Calculated spectral variations and light curves show distinct
behavior, depending on the nature of the emitting region;
time variation of a few months with strong
wavelength  dependence is expected in the case of an
optically thick disk ({\it standard disk}), 
while an optically thin disk ({\it advection-dominated disk}) will
produce shorter, nearly wavelength-independent variation. In the
case of an associated circumnuclear starburst region 
much slower variations (over a year) will be superposed 
on the shorter variations caused by microlensing of the disk. 


\end{abstract}

\section{Introduction}

There has been long discussion regarding the central
structure of active galactic nuclei (AGN).
Generally, it is believed that there is a supermassive
black hole and a surrounding accretion disk.
However, it is quite difficult to obtain direct
information about the center of AGN, 
the accretion disk, because the disk size is far too 
small to resolve.
As for cosmological objects, e.g., quasars, 
the (angular diameter) distance is of the order of 
$\sim 1 {\rm Gpc} \sim 3 \times 10^{27}$cm.
If a black hole placed at the center 
has a mass of $\sim 10^{8}M_{\odot}$, the disk size 
(say $\sim 1000r_{\rm g}$, where $r_{\rm g}$ is Schwarzschild radius) 
will be $\sim 3 \times 10^{16}$ cm.
Hence, apparent disk size ($\theta_{\rm disk}$) is 
only $\sim 10^{-11} {\rm radian} \sim 1 \mu{\rm as}$.

What is usually done is to observe the spectra, which are
the sum of local spectra over the entire disk,
and to make a fitting to the observed spectra by
theoretically integrated spectra.  However, there remain
ambiguities as to the spatial emissivity distribution.
The standard picture is that the so-called UV bump is
attributed to blackbody radiation from the standard-type, 
optically-thick disks (i.e., Shakura \& Sunyaev 1973).  
Such a fundamental
belief may be reconsidered by the recent HST observations
that show no big blue bumps (e.g., Zheng et al. 1997).

Instead, microlensing can be used as a `gravitational telescope' 
to resolve the disk structure (see Blandford \& Hogg 1995; originally
proposed by Chang \& Refsdal 1984),
since a typical Einstein-ring radius of a stellar mass lens is very small; 
e.g., $\theta_{\rm E} \sim 10^{-11} {\rm radian}$ 
for cosmological distant sources. 
With such a gravitational telescope we may be able to 
obtain information about the disk emergent spectra 
as a function of the distance from the central black hole.

When a point source object is lensed, 
its luminosity is amplified without any color change. 
This feature is called `achromaticity' and is discussed 
in many literature, initiated by Paczy\'nski 
(1986; for a review see Narayan \& Bartelmann 1996).
But, in some actual situations, we cannot treat a source as a
point and the effect of a finite source size should be taken into account. 
Grieger, Kayser \& Refsdal (1988) examined such finite size effects. 
They simulated microlensing light curves for two different
source profiles: stellar like and accretion disk like. 
The light curves show distinct shapes from those of point
source calculations. 
However, they only considered monochromatic intensity variation.
Later, Wambsganss \& Paczy\'nski (1991) pointed out that
if the finite size source has inhomogeneous structure 
so that each part produces a different spectrum,  
lensing light curves have wavelength dependence; 
i.e., chromatic features in microlensing events should appear.
This chromatic effect has been investigated by several authors 
who calculated microlensing light curves of AGN accretion disk
based on simple disk models (e.g., Rauch \& Blandford 1991; 
Jaroszy\'nski, Wambsganss \& Paczy\'nski 1992).

The purpose of this paper is to obtain some criteria for 
distinguishing disk physics based on more realistic disk models,  
we adopt the optically-thick, standard-type disk and 
the most successful, optically thin disk model, 
the advection-dominated accretion flow 
(ADAF; Narayan \& Yi 1995; Abramowicz et al. 1995).  
So far nobody (except Yonehara et al. 1998) has yet calculated X-ray
or radio emission in the microlens events, 
due probably to the lack of reliable disk models. 
However, such calculations are of great importance 
since X-ray and radio emission seems to originate 
from the vicinity of a putative black hole.
In section 2 we describe our methods of simulation. 
In section 3 our results are displayed. 
Section 4 is devoted to discussion.

\section{Method of the simulation}

\subsection{Accretion disk models}

First, we describe the adopted accretion disk models.
We consider two representative types of disks:

\noindent{1. \bf Optically thick disk:}

\noindent{The}
first case we consider is the geometrically-thin and
optically-thick standard disk (Shakura \& Sunyaev 1973).
Since we are concerned with disk structure 
on scales ranging from a few tens to a few thousands of $r_{\rm g}$,
we ignore relativistic effects, 
such as the gravitational redshift by the black hole, for simplicity.
The temperature distribution of the optically thick standard disk
is then given by (Shakura \& Sunyaev 1973) 
\begin{eqnarray}
T(r) = 2.2 \times 10^5 \left( \frac{\dot{M}}{10^{26}
 {\rm g \ s^{-1}}} \right)^{1/4} \left( \frac{M}{10^8 M_{\odot}}
\right)^{1/4}  \nonumber \\
\times \left( \frac{r}{10^{14} {\rm cm}} \right)^{-3/4}
\left[1 - \left( \frac{r_{\rm in}}{r} \right)^{1/2} \right]^{1/4} {\rm K},
\end{eqnarray}
where $\dot{M}$ is the mass accretion rate and
$M$ is the black hole mass.
The inner edge of the disk, $r_{\rm in}$, is set to be 
$r_{\rm in}=3r_{\rm g}$,
the radius of the marginally stable last circular orbit around
a non-rotating black hole.

Since the disk is optically thick, we can assume blackbody radiation.
Inserting the temperature profile $T(r)$ into 
\begin{equation}
\Delta L(\nu,r) = 2 \cdot 4 \pi B_{\nu} \left[ T(r) \right] 
                   \Delta S \cos \alpha,
\label{sandsl}
\end{equation}
we can calculate the luminosity of a part of the disk 
with the surface element, $\Delta S = \Delta r \cdot r \Delta \phi$, 
at a radius $r$ and a frequency $\nu$, 
where $B_{\nu}$ is a Planck function 
(e.g., see eq.(1.51) of Rybicki \& Lightman 1979), 
and $\alpha$ is the inclination angle of the disk.
For simplicity, here, we assume the disk is face-on ($\alpha = 0$).
It might be noted that the radiation spectra do not depend on
the viscosity in this standard-type disks, 
since the total emissivity and 
the effective temperature are solely determined 
by the energy balance between the radiative cooling and viscous heating 
(which originates from release of gravitational energy).

\noindent{2. \bf Optically thin disk}

\noindent{The}
second model is the optically thin version 
of the advection dominated accretion flow (ADAF).
We used the spectrum calculated by Manmoto, Mineshige, \& Kusunose (1997)
and derive luminosity $\Delta L(\nu,r)$ in the same way 
as we did for an optically thick disk: 
\begin{equation}
\Delta L(\nu,r)= 2 \cdot \epsilon_{\nu}(r) \Delta S,
\label{adafl}
\end{equation}
where $\epsilon_{\nu}$ is surface emissivity 
which include the effect of inclination angle 
(details are given in Manmoto et al. 1997).
In this paper, we also set $\alpha = 0$ (face-on) 
same as the case of optically thick disk.
Here, included are synchrotron, bremsstrahlung, and inverse
Compton scattering of soft photons created by the former two processes
(see Narayan \& Yi 1995 for more details).
Optical flux is mainly due to Comptonization of synchrotron photons.

In this sort of disks (or flows), 
radiative cooling is inefficient because of very low density.
As a result, accreting matter falls into a central object, 
hardly losing its internal energy 
(which is converted from its gravitational energy 
through the action of viscosity) in a form of radiation.  
Hence, although the total disk luminosity is less than 
that of the standard one for the same mass-flow rate,
the disk can be significantly hotter,
with electron temperatures being of the order of $10^9$K or more,
thus producing high energy (X-$\gamma$ ray) photons.
Photon energy is thus widely spread over large frequency ranges.
Since emission inside the last circular orbit (at $3r_{\rm g}$)
is not totally negligible in this case, we solve the flow structure
until the event horizon ($r_{\rm g}$) passing through the transonic point,
and consider radiation from the entire region outside the horizon,
although the contribution from the innermost part is not dominant.
Unlike the optically thick case, the viscosity explicitly affects
the emissivity in this case, since radiative cooling 
depends on density and is no longer balanced with viscous heating 
(and with gravitational energy release).
In the present study, we assign the viscosity parameter to be $\alpha=0.1$.

\subsection{Microlensing events}

Although single lens approximation is not always appropriate 
to the microlensing events, we, here, use this simple method 
because of making the situation clear 
(for extend source effect of microlensing by single lens object, 
 e.g., Bontz 1979).

Generally, the microlensed images are split into two (or more) images.  
When the separation between these images is too small to resolve, however,
the total magnitudes of all the split images vary due to this microlensing
caused by the relative motion of lens and source.

The properties of microlensing have been extensively examined 
for simple cases where both the lens and source 
are regarded as being points (e.g., Paczy\'nski 1986). 
In our simulation, we include an effect of occultation by a lens object 
which is recently included the simulation of microlense 
light curves by Bromley (1996) for reality.
In other words, we, here, consider that the radius of the lens object
($R_{\rm lens}$) is same as the sun, i.e.,
$R_{\rm lens} = R_{\odot} = 6.96 \times 10^{10} ~{\rm cm}$,
and calculate image positions of the all cell.
If the image position of a cell from the lens object is
smaller than the radius of lens object,
we exclude a contribution from that image.
Therefore, the total amplification (or magnification) factor $A$ 
for microlensing is exactly expressed as 
\begin{equation}
A(u) = \left\{ 
 \begin{array}{@{\,}ll} 
  0 & \left( \theta_{{\rm image,}+} <   
    \frac{R_{\rm lens}}{D_{\rm ol}} \right) \\
  \frac{u^2 + 2}{2u \left( u^2+4 \right)^{1/2}} + \frac{1}{2}  & 
    \left( \theta_{{\rm image,}-} < \frac{R_{\rm lens}}{D_{\rm ol}} 
    \le \theta_{{\rm image,}+} \right) \\
  \frac{u^2+2}{u \left( u^2+4 \right) ^{1/2}} & ({\rm otherwise}),
 \end{array}
\right.
\label{pointampli}
\end{equation}
where $u$ corresponds to the angular separation between lens and source
in the unit of the Einstein-ring radius and 
always greater than or equal to zero, i.e., $u \ge 0$, 
and $\theta_{{\rm image,}\pm}$ shows that the image positions of 
two microlensed images from a lens object in the unit of radian, 
i.e., 
\begin{equation}
\theta_{{\rm image,}\pm} = \frac{1}{2} \theta_{\rm E} 
  \left[ u - \left( u^2 + 4 \right)^{1/2} \right]
\end{equation} 
This function as shown in eq.~\ref{pointampli} 
is a monotonically increasing function of $u$; 
$A(u) \simeq 1.34$, $\sim 10$, $\sim 100$ for 
$u = 1.0$, $0.1$, and $0.01$.

Here, the Einstein-ring radius is
\begin{equation}
\theta_{\rm E} = \left(\frac{4GM_{\rm lens}}{c^2} 
                       \frac{D_{\rm ls}}{D_{\rm os} D_{\rm ol}}\right)^{1/2}
\end{equation}
where $M_{\rm lens}$ is the mass of a lensing star, 
$c$ is the speed of light, $G$ is the gravitational
constant, and $D_{\rm ls}, D_{\rm os}$, and $D_{\rm ol}$ 
denote the angular diameter distances from lens to source, 
from observer to source, and from observer to lens,  
respectively (Paczy\'nski 1986).
Adopting an appropriate cosmological model, 
we can calculate these angular diameter distances in terms
of redshifts $z_{\rm ls}$ (from lens to source), 
$z_{\rm os}$ (from observer to source), and $z_{\rm ol}$ 
(from observer to lens).
In the present study, we assume the Einstein-de Sitter Universe, 
in which we have
\begin{equation}
D_{\rm x} = \frac{2c}{H_{\rm 0}} \frac{1}{1+z_{\rm x}} 
\left[ 1- \frac{1}{ \left( 1+z_{\rm x} \right) ^{1/2}} \right] 
\end{equation}
(e.g., Padmanabhan 1993), where $H_{\rm 0}$ is Hubble's constant, 
and the subscript `${\rm x}$' stands for 
`${\rm ls}$', `${\rm os}$', or `${\rm ol}$'.

\begin{figure}
\resizebox{\hsize}{!}{\includegraphics{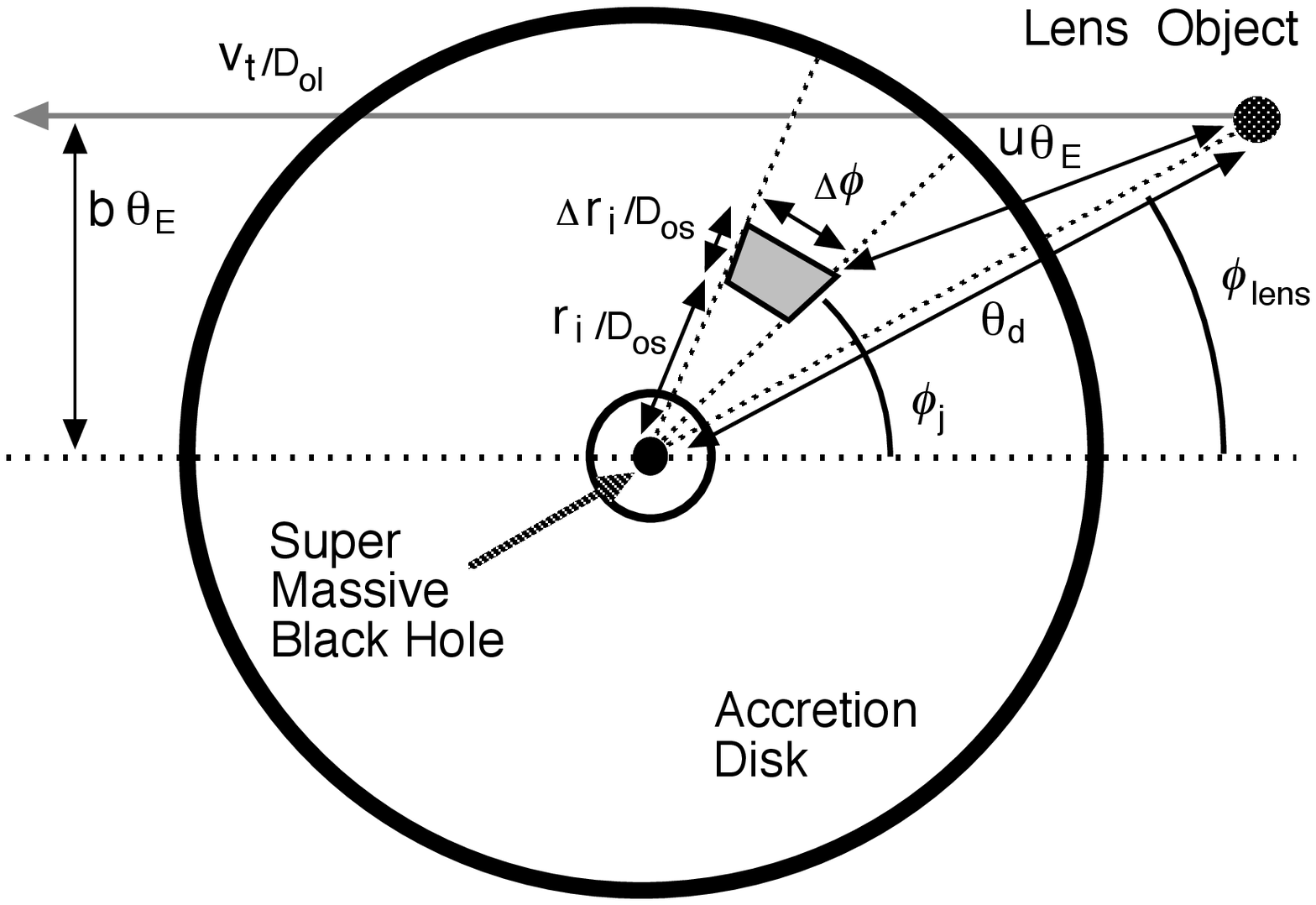}}
\caption[]{Schematic view of calculation for the separation between 
each cell and lens. }
\label{schv1}
\end{figure}

Since we now consider microlensing effects on an extended source, 
we must integrate the magnification effects over the entire source plane.
We thus, as a next step, need to calculate the angular separation $u$ 
between a part of the source in question and the lens center 
(see figure 1), from which the amplification factor, $A(u)$, can be found.
In the present study, we first divide an accretion disk
plane into azimuthal ($\phi$) and radial ($r$) elements.
The azimuthal coordinate is equally
divided into $\sim 1000$ segments ($\phi_j$), while the 
logarithmically scaled radial coordinate is also equally 
divided into $\sim 1000$ segments ($r_i$).
We then calculate the observed flux at frequency 
$\nu$ from a cell between $r_i$ and $r_{i+1}$ in the radial 
direction and between $\phi_j$ and $\phi_{j+1}$ in the 
azimuthal direction.
Noting that photons emitted by the source
at a frequency $\nu$ are observed at the frequency $\nu/(1+z_{\rm os})$,
we write the observed flux as 
\begin{equation}
 \Delta F_{\rm obs}(\nu;r_i,\phi_j) 
   \simeq A(u) \Delta F_{\rm obs,0}
    \left[\nu(1+z_{\rm os}),r_i,\phi_j \right], 
\end{equation}
where $A(u)$ is the amplification factor given by equation (4).
Moreover, $u$ is explicitly given by
\begin{eqnarray}
u &=& \left[ \left( \frac{r_i}{D_{\rm os}}\cos\phi 
        - \theta_{\rm d}\cos\phi_{\rm lens} \right)^2 \right. \nonumber \\
  &+& \left. \left( \frac{r_i}{D_{\rm os}}\sin\phi 
        - \theta_{\rm d}\sin\phi_{\rm lens}\cos \alpha \right)^2 
          \right]^{1/2} \theta_{\rm E}^{-1},
\end{eqnarray}
where $\alpha$ is the inclination angle of an accretion disk 
(in this paper, as indicated in \S 2.1, we set $\alpha = 0$), 
and $\theta_{\rm d}$ and $\phi_{\rm lens}$ represent the radial
and the azimuthal positions of the lens in units of radians, respectively.
Schematic view of the calculated disk is shown in figure~\ref{schv1}.
The intrinsic flux in the absence of a microlensing event, 
$\Delta F_{\rm obs, 0} (\nu, r)$, 
is calculated from $\Delta L(\nu,r)$ according to 
\begin{equation}
\Delta F_{\rm obs, 0} (\nu ,r) = 
  \frac{\Delta L(\nu ,r)}{4 \pi D_{\rm os}^2 (1+z_{\rm os})^3},
\end{equation}
and $\Delta L(\nu,r)$ depends on the disk model 
(see eq.~(\ref{sandsl}) and (\ref{adafl})).
In the present study, since we neglected the relativistic effect  
(e.g., beaming, gravitational redshift etc.), 
$\Delta L(\nu;r,\phi)$ has no azimuthal dependence.  

Summing up $\Delta F_{\rm obs}(\nu;r,\phi)$
over the entire disk plane from the inner boundary 
($r_{\rm in}$) to the outer boundary ($r_{\rm out}$), 
we obtain the total observed flux at frequency $\nu$, 
or the spectrum of the microlensed accretion disk.

\section{Results of the calculations}

\subsection{Model parameters}

For clarity, we consider a lensed quasar, 
Q~2237+0305 (the Einstein cross, e.g., Huchra et al. 1985). 
Irwin et al. (1989) reported an increase of the apparent luminosity 
of image A by $\sim 0.5$mag on a timescale of a few months, 
which was nicely reproduced by a model of microlensing 
(Wambsganss, Paczy\'nski, \& Schneider 1990).
At present, at least five microlens events have been reported so far
on the split images (Irwin et al. 1989; Corrigan et al. 1991;
Houde \& Racine 1994).
Although the single lens treatment for Q~2237+0305 
may be a poor approximation (see discussion), 
we just take this object as one test case
and derive quite general conclusions regarding 
the generic observational features of microlensed accretion disks.

This object is macrolensed by a foreground galaxy, and the
source is split into four (or five) images.
Now, we suppose that one of the macrolensed image is further microlensed 
by a star in the foreground galaxy causing the macrolensing. 
Although the macrolensed images are actually affected by
two effects, so called, `convergence' and `shear', we here
neglect both for simplicity and thus assume
that the macrolensed image is neither amplified nor distorted. 
The redshift parameters are
\begin{equation}
z_{\rm os} = 1.675, \ z_{\rm ol} = 0.039 
\end{equation}
(cf. Irwin et al. 1989), 
and substitute them into the relation 
\begin{equation}
1 + z_{\rm ls} = \frac{1 + z_{\rm os}}{1 + z_{\rm ol}}
\end{equation}
which can be obtained from flux conservation of radiation 
(e.g., Padmanabhan 1993), we get  
\begin{equation}
z_{\rm ls} = 1.575.
\end{equation}
As for Hubble's constant, we adopt
the value obtained by Kundi\'c et al.(1997), 
$H_{\rm 0} \sim 60 {\rm km \ s^{-1} Mpc^{-1}}$.
The adopted disk parameters are
the outer edge of the disk, $r_{\rm out} = 10^3r_{\rm g}$, 
the mass of the central black hole, $M = 10^8 M_{\odot}$,
and the mass-flow rates,
$\dot{M} =10^{26}$ g s$^{-1}$ for the standard disk and
$\dot{M} =10^{22}$ g s$^{-1}$ for the optically thin disk.
We take different mass-flow rates for two cases because of
calculation convenience.  This is not, fortunately, a serious problem,
since it has been demonstrated that the spectral shape of the ADAF
is not very sensitive to $\dot M$,
although the overall flux level does change.
We will not discuss the absolute flux in the following discussion,
and focus on the relative spectral shape and relative flux variation.
For other basic parameters characterizing the disks, see \S 2.1.

For simplicity, we assume that the accretion disk is face-on;
i.e. $i = 0$, to the observer. Actually, based upon a grand unified
paradigm of AGNs, quasars are not far from face-on views. 
We also assume the mass of a lensing object to be 
$M_{\rm lens}=1.0M_{\odot}$ or $0.1M_{\odot}$.
The remaining variable is the angular separation ($u$) between the lens
and a part of the source in question.

Adopting these parameters, apparent angular accretion disk size 
($\theta_{\rm disk}$) and Einstein ring radius ($\theta_{\rm E}$) 
is $\theta_{\rm disk} \sim 6.8 \times 10^{-12} {\rm rad}$ and 
$\theta_{\rm E} \sim 3.3 \times 10^{-11} {\rm rad}$.
These values are comparable and we, here,  can use microlensing event 
as effective `gravitational telescope' (Blandford \& Hogg 1995).

\subsection{Spectral variation}

\begin{figure}
\resizebox{\hsize}{!}{\includegraphics{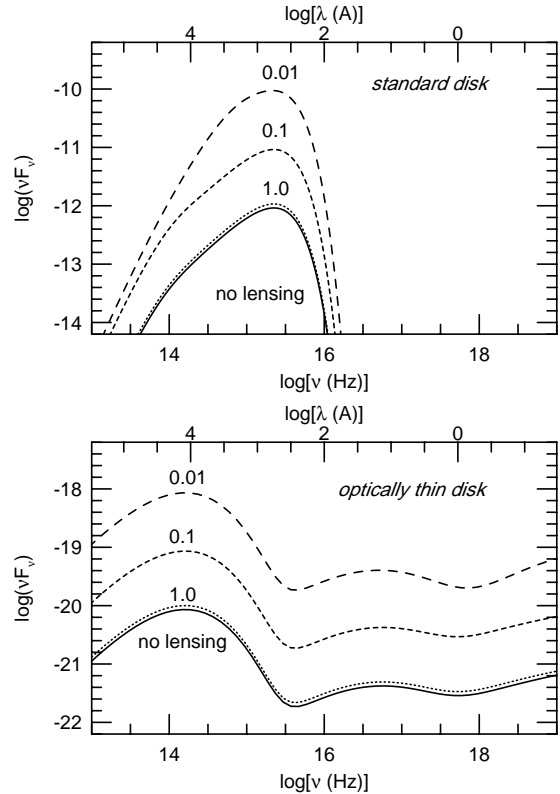}}
\caption[]{Microlensed spectrum of a standard accretion
disk (upper) and an optically thin accretion flow (lower).
In each panel, the microlensed spectra are plotted for
angular separations between the disk center and the lensing star to be
$u=1.0$ (dotted line), $0.1$ (dashed line), and $0.01$ (long dashed line),
together with the one in the absence of microlensing (solid line).
The lensing mass is $M_{\rm lens} = 1.0 M_{\odot}$.}
\label{fig1}
\end{figure}

First, we calculated the microlensed spectra of each of two types
of accretion disks for the angular separation between the lens and the source 
of $u=1.0$, 0.1, and 0.01, respectively.
The results are shown in figure~\ref{fig1} for the standard disk 
(in the upper panel) and for the optically thin disk (in the lower panel), 
respectively.
For the former disk not only the disk brightening, 
but also substantial spectral deformation is produced by microlensing,
when the angular separation is small
(Rauch \& Blandford 1991; Jaroszy\'nski, Wambsganss \& Paczy\'nski 1992).
The smaller the angular separation is, the larger becomes the modification
of the observed spectrum.
In other words, the amplification factor depends on the 
frequency (wavelength) of the emitted photons. 
This gives rise to frequency-dependent, microlensing light
curves (see \S 3.3).
As for the optically thin disk, in contrast, there will not
be large spectral modification by microlensing, but the flux is
amplified over wide frequency ranges.

\begin{figure}
\resizebox{\hsize}{!}{\includegraphics{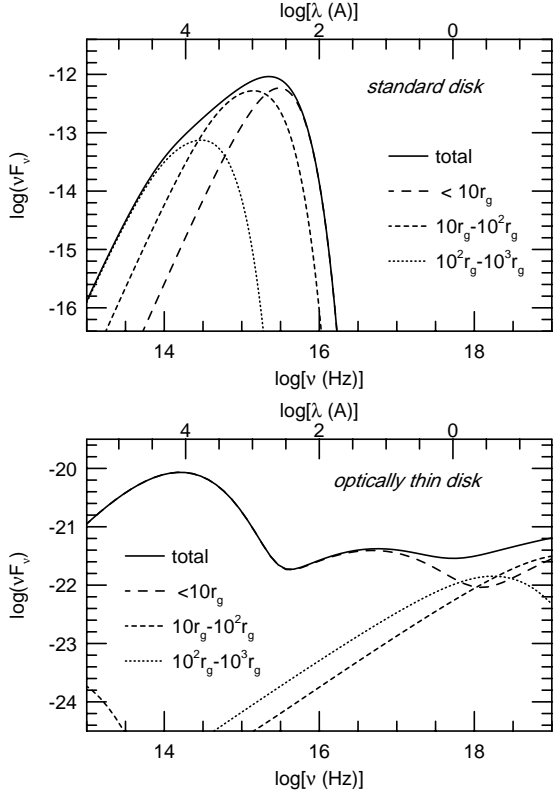}}
\caption[]{Contribution to the total integrated spectra
from different parts of disks for the cases of a standard (upper)
and optically thin (lower) disk, respectively.
Each figure displays the contributions from
the inner region of $r \le 10r_{\rm g}$ (long dashed line),
the intermediate region of $10r_{\rm g} \sim 100r_{\rm g}$ (dashed line),
and the outer region of $10^2r_{\rm g} \sim 10^3r_{\rm g}$ (dotted line).}
\label{fig2}
\end{figure}

To understand such spectral properties, we divide the disk
plane into three parts; $r \le 10r_{\rm g}$,
$10r_{\rm g} < r < 100r_{\rm g}$, and $10^2r_{\rm g} < r <10^3r_{\rm g}$,
and plot the spectrum from each part of the disk, 
together with the integrated spectra, in figure~\ref{fig2}.

In the case of the optically thick accretion disk, photons
emitted from different radii have different dominant 
frequencies because the local temperature of the disk has a
strong radial dependence, $T \propto r^{-3/4}$ (Eq. [1]).
Note that without lensing the total spectrum has a peak
at $\nu \sim 10^{15.3}$ Hz, but its low-frequency part has a smaller slope
($\nu F_{\nu} \propto \nu^{4/3}$) than that of blackbody radiation
($\nu F_{\nu} \propto \nu^3$), thus a shoulder like feature extending
to $\nu \sim 10^{14}$ Hz $(\sim 30000 \AA)$ being formed.
This is because the outer cooler portions also contribute to the spectrum.
For $u=0.01$, therefore, the emission only from the inner,
hottest part is strongly magnified, 
much strengthened a peak at $\nu \sim 10^{15.3}$ Hz $(\sim 1500 \AA)$ 
and a shoulder like feature at $\nu \sim 10^{14}$ Hz $(\sim 30000 \AA)$ 
appears to be weakened.
Consequently, microlensed spectrum became a form with one sharp peak.

\begin{figure}
\resizebox{\hsize}{!}{\includegraphics{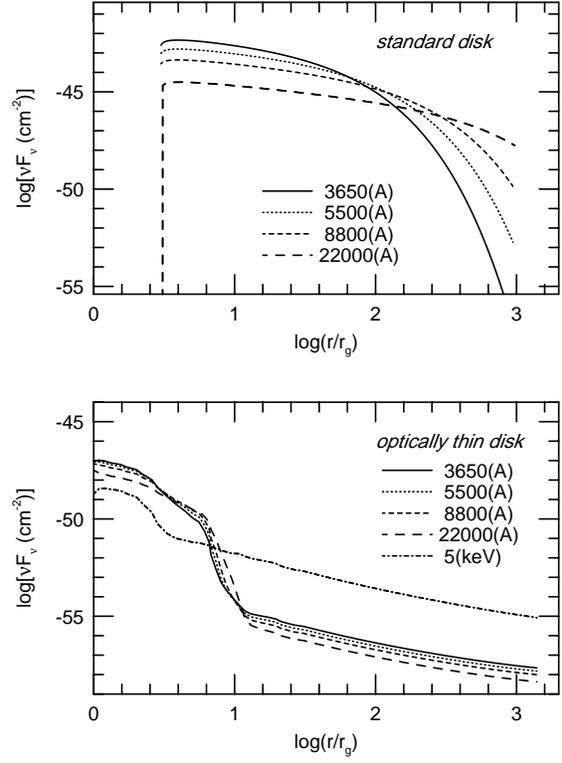}}
\caption[]{The radial distribution of the emergent flux with several
wavelengths for the case of a standard-type disk (upper) and
an optically thin disk (lower), respectively.
The adopted bands are
U-band (3650 \AA, by the solid line),
V-band (5500 \AA, by the dotted line),
I-band (8800 \AA, by dashed line),
K-band (22000 \AA, by long dashed line), and
X-ray  (5 keV, by the dot-dashed line). }
\label{fig3}
\end{figure}

To help understanding such situation, we also plot
the radial distribution of emitted photons with several wavelengths
in figure~\ref{fig3}.
The upper panel of figure~\ref{fig3} shows rather gradual change of each flux
over a wide spatial range.  
This reflects the fact that viscous heating and radiative cooling
are balanced in the standard disk.  Since the potential energy
only gradually changes with the radius, so is the viscous heating rate
and is the radiative cooling rate.  
Moreover,
the lower a photon frequency is (or the longer the wavelength is),
the wider becomes the parts of the disk which generate photons.

In the optically-thin accretion disk, on the other hand,  
the optical-UV flux is totally dominated by that from the inner region 
($r \le 10r_{\rm g}$).  
The lower panel of figure~\ref{fig3} clearly shows 
that almost 100\% of optical flux originates from the region 
inside $10r_{\rm g}$, and that the contribution from 
the outer parts is smaller by several orders.
This is possible in ADAF, since radiative cooling is no longer 
directly related to the shape of the gravitational potential well 
(which has a rather smooth profile).
The reason why optical flux is produced in a rather restricted region
inside $10r_{\rm g}$ is due to enhanced synchrotron emission
in the vicinity of the black hole, where density and pressure
(and thus magnetic pressure) are likely to be at maximum within the flow.
Nearly irrespective of the $u$ value, as a consequence,
it is always the innermost region (within 10 $r_{\rm g}$) 
that dominates the entire disk spectrum.
The amplification of the total emission just reflects
that of the emission from the innermost parts.
Therefore, the overall flux will be
amplified by lensing without large spectral changes.

Figure~\ref{fig3} also shows that X-rays are produced over a wider range; 
the contribution from the outer part at $100 - 1000 r_{\rm g}$
is not negligible at frequency of $f \sim 10^{18}$ (a few keV).
This is because of a bremsstrahlung emission from 
high-temperature electrons (with $T_{\rm e} \sim 10^9 K$ being 
maintained at $r \lsim 100 r_{\rm g}$, see Manmoto et al. 1997).
We can thus predict that small frequency dependence will be seen in
X-ray ranges in this model.

Such distinct spectral behaviors give rise to 
different light curves of the two disk models (see below).

\subsection{Light curves}

\begin{figure}
\resizebox{\hsize}{!}{\includegraphics{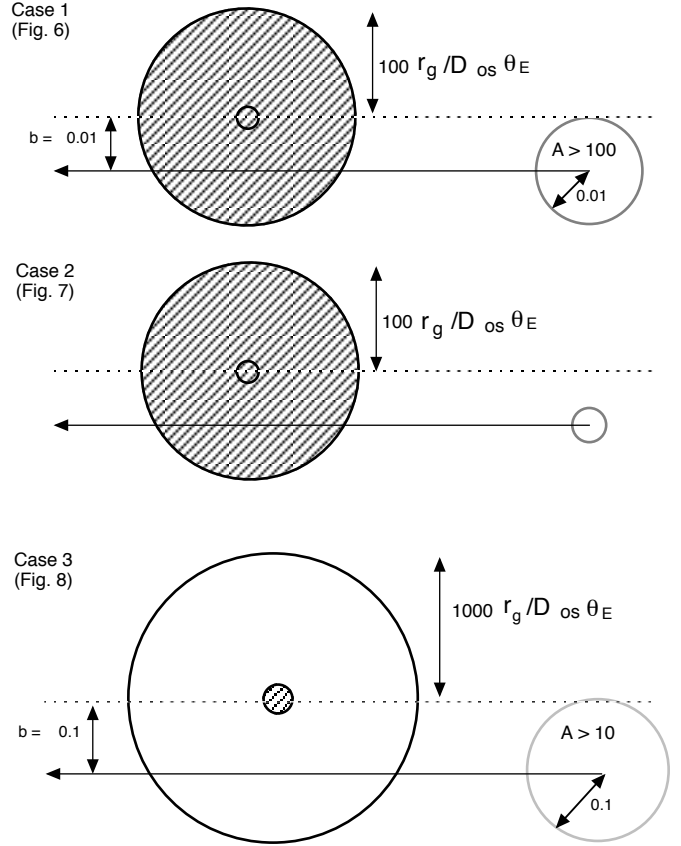}}
\caption[]{Schematic view of the microlensing events which we have calculated.
These figures show the relative sizes of the accretion disk, 
impact parameter, tracks which lens object passes, 
and the Einstein-ring radius. 
In each panel, shaded regions correspond to 
the disk with a radius of $100r_{\rm g}$.
Top panel (Case 1) shows microlensing event of
$(b, M_{\rm lens})=(0.01,1.0 M_{\odot})$.
Left, thick outer and inner circles corresponds to
the radius of $100r_{\rm g}$ and $10r_{\rm g}$, respectively.
A right circle corresponds to $\sim 0.01\theta_{\rm E}$.
Inside this circle, the amplification factor caused by microlensing
is more than $100$.
Middle panel (Case 2) shows microlensing event of
$(b,M_{\rm lens})=[0.01(M_{\odot}/M_{\rm lens})^{1/2}, 0.1M_{\odot})]$.
The types of left circles are the same as those in the top panel.
Bottom panel (Case 3) shows microlensing event of
$(b,M_{\rm lens})=(0.1, 1.0M_{\odot})$.
Left, thick outer and inner circles corresponds to
$1000r_{\rm g}$ and $100r_{\rm g}$ of accretion disk.
A right circle corresponds to $\sim 0.1\theta_{\rm E}$, 
amplification factor caused by microlensing
is more than $\sim 10$.}
\label{disks}
\end{figure}

Using the results of the previous subsection, 
we now calculate the light curve during the microlensing events 
(see fig.~\ref{schv1}).
We continuously change the
angular separation between the lens and the center of the 
accretion disk ($\theta _{\rm d}$) according to  
\begin{equation}
\theta _{\rm d} = \theta_{\rm d}(t) = \theta_{\rm E} \left[ b^2 + \left( 
\frac{v_{\rm t} t}{D_{\rm ol} \theta_{\rm E}} \right)^2 \right] ^{1/2},
\end{equation}
and calculate the flux at each frequency, 
where $b$ is the impact parameter (in the unit of $\theta_{\rm E}$), 
$v_{\rm t}$ is the transverse velocity of the lens and $t$ is the time.
Note that $v_{\rm t}$ also includes the transverse velocity 
due to the peculiar motion of the foreground galaxy relative 
to the source and the observer.
We calculated three cases (see fig.~\ref{disks}).

\begin{figure}
\resizebox{\hsize}{!}{\includegraphics{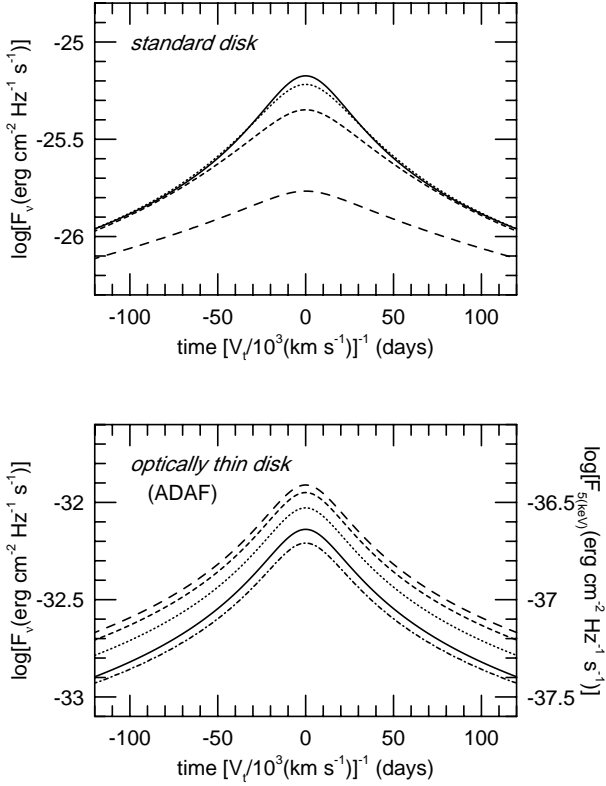}}
\caption[]{The microlensed light curves of the standard disk
(upper) and for the optically thin flow (lower).
The parameters are $(b, M_{\rm lens})=(0.01,1.0 M_{\odot})$ (Case 1).
The different type of lines correspond to the same as in figure~\ref{fig3}.}
\label{fig4}
\end{figure}

Figure~\ref{fig4} shows the microlensing light curves 
of optically thick (upper panel) and thin (lower panel) disks
for the case with $(b, M_{\rm lens}) = (0.01,1.0M_{\odot})$ (Case 1).
The light curve of the optically thin disk is still 
achromatic because of its flat temperature distribution, whereas 
the light curve of the optically thick one shows strong frequency dependence.

Figure~\ref{fig3} is again useful to understand this chromaticity 
and achromaticity.
Since flux distribution is similar among different optical fluxes
in the case of optically thin disks,
each optical flux is amplified in a similar way 
and thus the color does not change; i.e. achromatic feature appears. 
X-ray emitting region is wider than that of optical flux,
small chromaticity may appear when we compare optical and X-ray
light variations.

In an optically thick disk, conversely,
low-energy photons are created in a wider spatial range,
compared with high-energy photons.
If the microlensing event occurs, therefore,
low-energy emission will be the first to be amplified,
followed by the significant amplification of high-energy emission.
Note that each optical flux has the same level at the beginning
(at $-120$ day) 
in the upper panel of figure~\ref{fig4}, whereas a higher-energy optical flux
is greater than that of a lower-energy flux in the intrinsic spectra
in the absence of lensing
(see figure~\ref{fig2}).  This indicates that the amplification of a 
lower-energy flux has already been appreciable at $t = -120$ day.
This results in the color change; i.e. chromatic feature is produced.

\subsection{Parameter dependence}

\begin{figure}
\resizebox{\hsize}{!}{\includegraphics{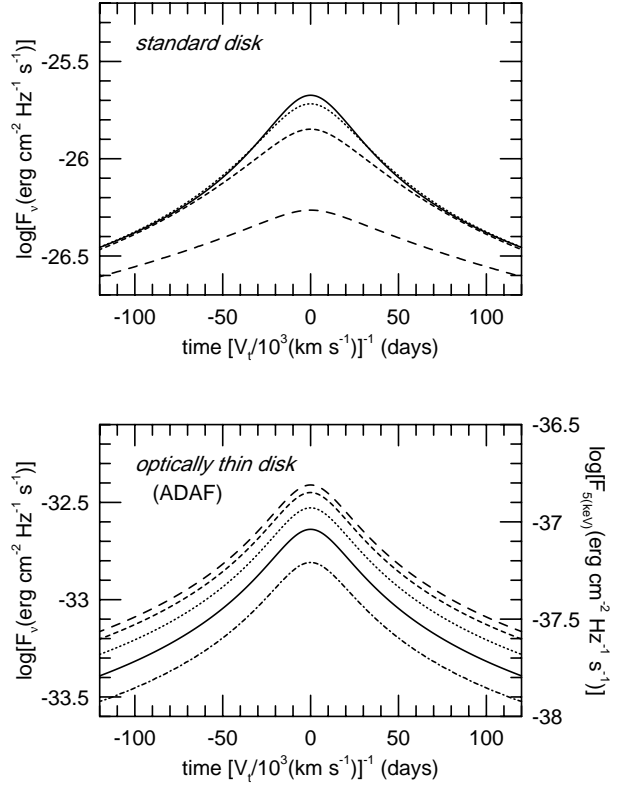}}
\caption[]{Same as figure~\ref{fig4} but for the case with 
smaller lensing mass;
$(b,M_{\rm lens})=[0.01(M_{\odot}/M_{\rm lens})^{1/2}, 0.1M_{\odot})]$
 (Case 2).}
\label{fig5}
\end{figure}

Figure~\ref{fig5} shows the light curves for the case 
with a smaller lensing mass by one order and
the same physical impact parameter ($b\theta_{\rm E}$) (Case 2).  
Namely,
$(b,M_{\rm lens}) = [0.01(M_{\odot}/M_{\rm lens})^{1/2},0.1M_{\odot}]$.
We see that if the lens mass is small, the peak magnification 
is reduced but the variability timescale is near the same at this range, 
while the chromaticity of the standard disk still remains.
This is more clearly shown in table 1.

\begin{table*}
\begin{center}
\begin{tabular}{l l l | l l l | l l} \hline
Case & disk model & band & 
$~ \log F_{\rm (200d)}$ & $~ \log F_{\rm (100d)}$ & 
$~ \log F_{\rm (0d)}$ & $\log [F_{\rm (100d)}/F_{\rm (200d)}]$ &
$\log [F_{\rm (0d)}/F_{\rm (100d)}]$ \\ \hline
Case 1 & standard & U & -26.179 & -25.883 & -25.174 & 0.296 & 0.709 \\ 
& & K & -26.288 & -26.061 & -25.766 & 0.227 & 0.305 \\ 
& ADAF & U & -33.113 & -32.820 & -32.139 & 0.293 & 0.681 \\ 
& & K & -32.885 & -32.592 & -31.910 & 0.293 & 0.682 \\ 
Case 2 & standard & U & -26.667 & -26.380 & -25.674 & 0.287 & 0.706 \\ 
& & K & -26.775 & -26.555 & -26.264 & 0.220 & 0.291 \\ 
& ADAF & U & -33.600 & -33.317 & -32.638 & 0.283 & 0.679 \\ 
& & K & -33.372 & -33.089 & -32.410 & 0.283 & 0.679 \\ 
Case 3 & standard & U & -26.340 & -26.247 & -26.203 & 0.093 & 0.044 \\ 
& & K & -26.446 & -26.352 & -26.311 & 0.094 & 0.041 \\ 
& ADAF & U & -33.273 & -33.180 & -33.137 & 0.093 & 0.043 \\ 
& & K & -33.045 & -32.952 & -32.909 & 0.093 & 0.043 \\ \hline
\end{tabular}
\end{center}
\caption{The calculated fluxes at $3650 \AA$ ({\it U}-band) 
and $22000 \AA$ ({\it K}-band) at different epochs for case 1,2, and 3. 
Second, and third columns indicate the adopted accretion disk models 
and the energy band. 
Fourth, fifth, and sixth columns show the observed fluxes 
at $t=200$ d (or $-200$ d), $100$ d (or $-100$ d), and $0$ d.
Seventh and eighth columns display
changes in the fluxes from $t=-200$ d to $t=-100$ d, and 
those from $t=-100$ d to $t=0$ d, respectively.
These two values are good indicators of the chromatic feature. }
\label{steepness}
\end{table*}

\begin{figure}
\resizebox{\hsize}{!}{\includegraphics{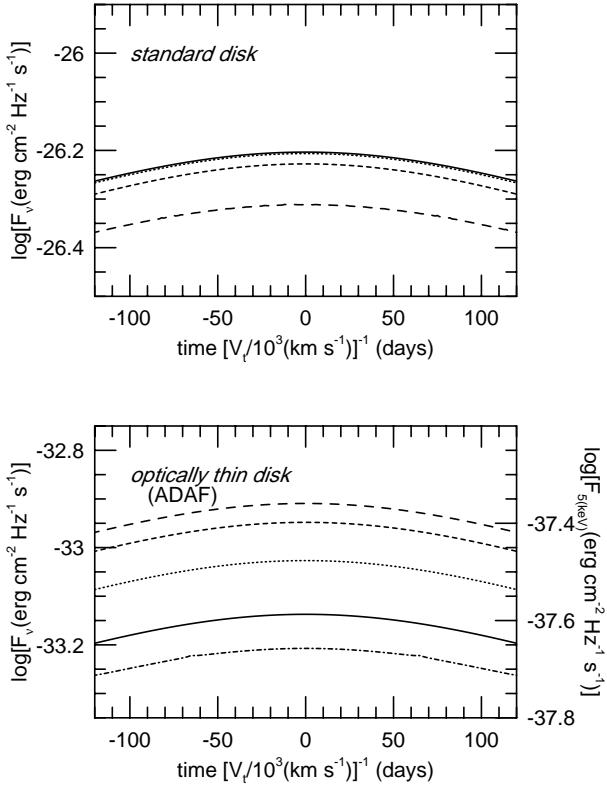}}
\caption[]{Same as figure~\ref{fig4} but for the case 
with larger impact parameter;
$(b,M_{\rm lens})=(0.1, 1.0M_{\odot})$ (Case 3).}
\label{fig6}
\end{figure}

Next, in figure~\ref{fig6}, we assign a larger impact parameter;
$(b,M_{\rm lens}) = (0.1,1.0M_{\odot})$ (Case 3).  
In this case,
not only the total amplification become small but also 
chromaticity of standard disks no longer holds.
Consequently, when we try to discriminate the disk structure, 
our success will depend not on the lens mass, 
but on whether the impact parameter is small or not.
We will discuss this issue later (in section 4).

In table~\ref{steepness}, we summarize characteristic values 
of these three light curves.
Achromatic and chromatic features are clear in this table. 

\begin{figure}
\resizebox{\hsize}{!}{\includegraphics{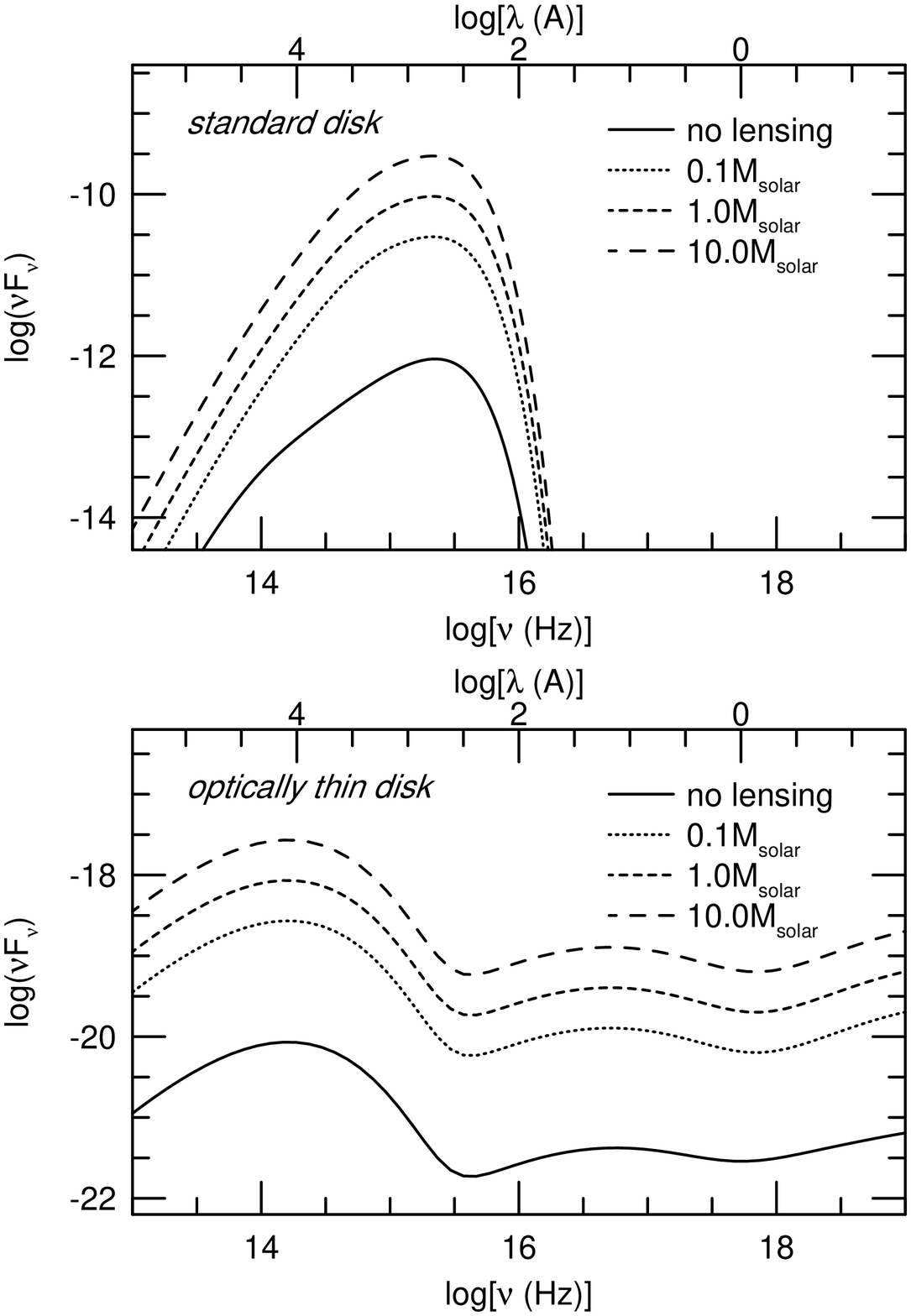}}
\caption[]{Dependence of the microlensed spectra of the
standard disk (upper) and the optically thin flow (lower)
on the lensing mass,
$M_{\rm lens} = 0.1 M_{\odot}$ (dotted line), $1.0 M_{\odot}$
(dashed line), and $10.0 M_{\odot}$ (long dashed line).
The impact parameters are $0.01(M_{\odot}/M_{\rm lens})^{1/2}$.
The spectrum without microlensing is also plotted by the solid line.}
\label{fig7}
\end{figure}

Finally, figure~\ref{fig7} shows the dependence of 
the spectrum on the lens mass. 
The smaller the lens mass is, the narrower 
the resulting shapes of the spectrum for a given flux value become.
This is because when the lens mass is small, 
the Einstein-ring radius is small
($\theta_{\rm E} \propto M_{\rm lens}^{1/2}$, see Eq. [5]),
so is the size of the region undergoing microlensing amplification.
This reduces the total light amplification.

\section{Discussion}

We have demonstrated how spectral behavior and 
microlensing light curves depend on the disk structure. 
We considered two representative disk models:
the standard-type disk emitting predominantly optical-UV photons,
and the optically thin, advection-dominated flow (ADAF)
producing a wide range of photons from radio to X-$\gamma$ rays.
Using the microlensing events, we can distinguish the different
emissivity distribution of AGN accretion disks and 
will be able to map them in details 
and reconstruct them from the light curves.

At the peak of the microlensing event, 
when the inner part of the accretion disk is largely amplified, 
distinctions between the two disk models are most evident
(especially when the impact parameter is small).
Thus, broad band photometry 
will be able to detect the color changes as shown in figure~\ref{fig4},
thereby revealing the structure of AGN accretion disks.

Fortunately, such observations do not require good time resolution; 
for example, once per night like the MACHO project 
(e.g., Alcock et al. 1998) is sufficient, 
as long as a lens is located far from the observer.
A typical timescale of microlensing event is characterized by 
the Einstein ring crossing time of a lens object expressed as 
\begin{eqnarray}
\frac{\theta_{\rm E}}{v_{\rm t}/D_{\rm ol}} \simeq & 6 & \times 10^6  
 \left( \frac{M_{\rm lens}}{M_{\odot}} \right) ^{1/2} \nonumber \\ 
& \times & \left( \frac{v_{\rm t}}{200 {\rm km~s}^{-1}} \right)^{-1} 
 \left( \frac{\tilde{D}}{10 kpc} \right)^{1/2} {\rm s}, 
\label{eq:times}
\end{eqnarray}
where $\tilde{D}$ means $D_{\rm ls}D_{\rm ol} / D_{\rm os}$.
For MACHO-like events when all the factors in the parentheses of 
eq.~(\ref{eq:times}) are of the order of unity, 
the event timescales become $\sim 0.2$ yr.
For the cosmological quasar case, on the other hand,  
we find roughly $\tilde{D} \sim 1{\rm Gpc} = 10^{5} \times 10{\rm kpc}$, 
and thus, the third parenthesis became large by factor $\sim 300$, 
yielding the event timescales being up to $\sim 60$ yr.
Thus, in principle, a few years observation is sufficient to 
obtain the information about central disk structure. 
Practically, reconstruction from microlensing light curve 
that is so-called `inverse problem' is not so easy,   
because, this requires a lot more to be done, 
e.g., take some (noisy, not perfectly sampled) light curve in a few filters. 
The source profile deconvolved by the observed light curve 
may comprise large errors, 
if the light curve sampling interval is 
much longer than the event duration timescale,
or if the accuracy of the observed magnitude (i.e., flux from the source) 
is much poorer, as stated before by Grieger, Kayser, \& Schramm (1991).
Therefore, we need good time resolution not for 
discriminating accretion disk model, 
but for mapping detailed accretion disk.
This reconstruction of accretion disk structure may be done 
in our future work.

\begin{figure*}
\resizebox{\hsize}{!}{\includegraphics{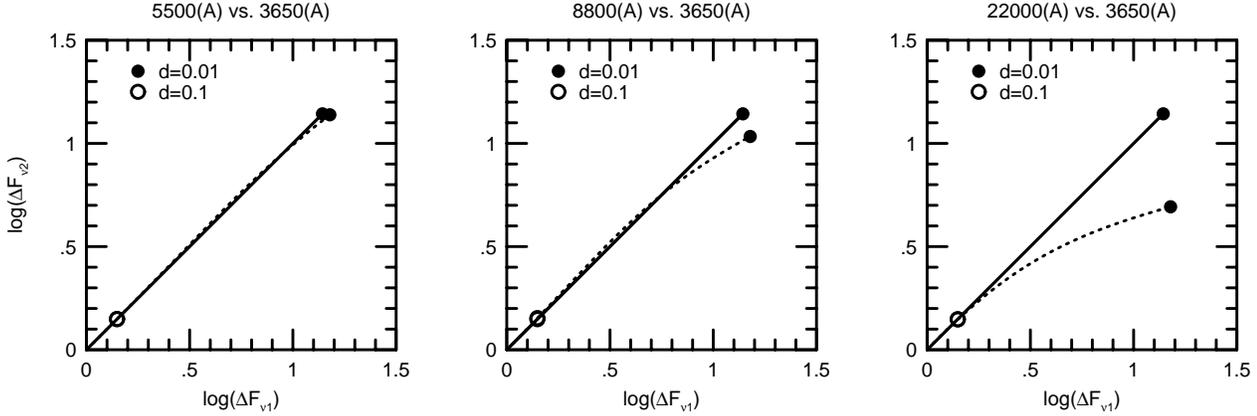}}
\caption[]{Differences in the amplification factors 
at two different frequencies
for the case of the standard disk (by the dotted line)
and the optically thin flow (by the solid line), respectively.
The adopted two frequencies are from the left, 
the U-band and the V-band (left panels),
the U-band and the I-band (middle panels),
and the U-band and the K-band (right panels), respectively.
Open and filled circles represent the cases with $b = 0.1$ and
$b=0.01$, respectively.
The lensing mass is assumed to be $M_{\rm lens}=1.0M_{\odot}$.}
\label{fig8}
\end{figure*}

To summarize the distinct properties of the two types of disks,
we plot in figure~\ref{fig8}
differences in the amplification factors at two different frequencies.
Open and filled circles represent the cases with $b = 0.1$ and 
$b=0.01$, respectively.
The lensing mass is assumed to be $M_{\rm lens}=1.0M_{\odot}$.
As the angular separation between the lens and 
the disk center ($u$) decreases, $\Delta F$ increases 
(thus the point moves in the upper right direction in figure~\ref{fig8}).
Obviously, the color changes are more pronounced in the right panel
(between the U- and K-bands)
than in the left panel (between the U- and V-bands), and
for smaller impact parameters.
We can easily discriminate the disk structure 
by comparing the optical and IR fluxes if $b \lsim 0.03$.
Furthermore, as clearly seen in figure~\ref{fig8}, 
the wider the separation between 
two observed wavelengths (or frequencies) are, 
the more apparent differences between the light curves 
of the two disk models become. 
Owing to this fact, the wider bands are preferable for 
more effective disk mapping. 

As a first step to understand the microlensing phenomenon, 
we have considered microlensing by a single star in the present study.  
However, the microlensing events caused by a single lens,
especially the case in which differences between the light curves of 
the two disk models are apparent (i.e., $b \le 0.1$), 
seems not so frequent for the actual objects.
We thus need to evaluate the average time interval 
$\langle \Delta t \rangle$ between microlensing events with
an impact parameter smaller than $b$.
From Paczy\'nski (1986),
we find that $\langle \Delta t \rangle$ is proportional to $b^{-1}$.
If we regard the event that the impact parameter is $b \le 1.0$ 
as the microlensing event, one microlensing event 
which we are interested in will occur among 
$10 ~(b=0.1)$ to $100 ~(b=0.01)$ events.

For example, we, here, assume the case of Q~2237+0305,  
although, in this case, single-lens assumption 
is not very good for this object 
because of its relatively large optical depth for microlensing 
(e.g., $\tau = 0.4 \sim 0.7$ from Rix, Schneider, \& Bahcall 1992).
It is widely believed that microlensing does occur in Q~2237+0305
at a rate of at least one event per year, i.e.
$\langle \Delta t \rangle \sim 1 {\rm yr}$
(Corrigan et al. 1991; Houde \& Racine 1994).
Hence, the microlensing event which we would like to observe 
is expected to occur once per 
$10$ yr (for $b=0.1$) to $100$ yr (for $b=0.01$).

To conclude, microlensing events 
with such a small impact parameter may not be frequent, 
but once the microlensing with a small impact parameter occurs,
a large light amplification is expected.  
We have demonstrated that
an optically thick accretion disk exhibits more curious features in
its spectral shape than an optically thin accretion disk.
Differences between these two disk models are, essentially, 
caused by different energy balance equations.
In an optically thick accretion disk, generally, viscosity is small, 
accreting gas stays in the accretion disk for a much longer time
than the disk rotation periods, so that  
gravitational potential energy of accreting gas 
can be effectively transformed into radiation 
on its way to the central compact object  
(in the present case, a supermassive black hole).
In an optically thin accretion disk, on the other hands, 
since viscosity is high, accreting gas falls into 
the central compact object rapidly in timescale 
on the order of the rotation period, 
and gravitational potential energy of accreting gas is not 
so effectively transformed into radiation 
(e.g., Kato, Fukue, \& Mineshige 1998). 
Thus, most of gravitational potential energy of accreting gas 
is advected into the central compact object and is not radiated away.
These major discriminant physical processes in the accretion disks 
can be distinguished by our microlens diagnostic technique.

In this paper, we assume that accretion disks are face on to the observer.
In real situation, however, accretion disks may have 
non-zero inclination angles (i.e., $i \neq 0$). 
It will be complicated to quantitatively treat  
the dependence of inclination angles 
because of the increase in the number of parameters, 
such as the angle between the path of the lens object 
and the apparent semi-major axis caused by a non-zero inclination.
Qualitatively, non-zero inclination affects 
timescale of the microlensing event in such a way to  
reduce the apparent disk size by factor $\sim \cos i$.
This effect seems to be small except the case 
that the disks are nearly edge on (i.e., $i \sim 90^{\circ}$). 
This inclination effect will be systematically examined in a future work.

\begin{figure}
\resizebox{\hsize}{!}{\includegraphics{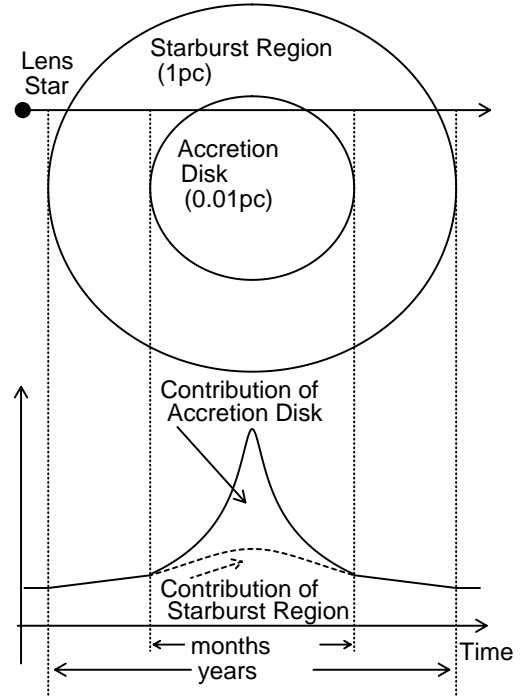}}
\caption[]{Schematic view of a microlensing event of an AGN disk and
circumnuclear starburst regions.
Upper figure shows the situation of microlensing of
accretion disk plus its circumnuclear starburst region.
Lower figure shows its expected light curve.}
\label{fig9}
\end{figure}
Finally, recent high-resolution observations, including those from {\it HST},
have revealed circumnuclear starburst regions in AGNs
(see Umemura, Fukue \& Mineshige 1997, and references therein).
If starburst regions exist around an accretion disk,
microlensing event of AGN will make an interesting light curve;
large-amplitude, short timescale variation (on a few months)
caused by the microlensing of the accretion disk will be superposed on
more gradual light variation (over a year) caused by the
microlensing of starburst region (see figure~\ref{fig9}).
As shown in figure~\ref{fig9}, the size of circumnuclear starburst regions 
is a tenth or hundredth as large as the central accretion disk.
Hence, the microlensing timescale of circumnuclear starburst region 
is one tenth or hundredth as long as that of the accretion disk and 
the total amplification of the circumnuclear starburst region is
smaller than that of the accretion disk from 
a rough estimate (see eq~(2.28) of Schneider, Ehlers, \& Falco (1992)).
Thus, contributions of these two compositions to 
the microlensing light curve will be able to discriminate,
and long-term monitoring may provide a possible tool
to elucidate the extension of circumnuclear starbursts 
although the presence of intrinsic variability 
tends to make such analyses complex.

\section{Summary}

The single-lens model as adopted in this paper is a good approximation, 
unless the apparent separations among lens stars are appreciably smaller 
than the radius of the Einstein ring in the source plane.
But, as we stated above, 
we should note that the images of Q2237+0305 could be subject to
microlensing by multiple stars of the foreground galaxy.
That is, we must consider, as a next step, 
the microlensing events caused by `caustics' which is produced
by multiple microlensing.
Such events may occur more frequently with shorter durations
and steeper magnification (Wambsganss \& Kundi\'c 1995).
Light curves by the caustics will show even more 
interesting features than the present case 
(Schneider \& Weiss 1986).
More realistic, analytically approximated `caustic' calculation 
has been done for the case of Q~2237+0305 (Yonehara et al. 1998).

This diagnostic method for the central region of AGN can be 
used in other sources (e.g., MG0414+0534, PG1115+080, 
and the so-called `clover leaf', H1413+117 etc.).
Future systematic and wide survey of quasars will discover other 
multiple quasars which our microlens diagnostic technique can apply. 

\begin{acknowledgements}
We thank Joachim Wambsgan{\ss} for valuable comments.
We acknowledge the referee for his/her helpful indications and 
precious comments.
One of the authors (A.Y.) also thanks Hideyuki Kamaya for his encouragement. 
This work was supported in part 
by Research Fellowships of the Japan Society for the
Promotion of Science for Young Scientists, 9852 (A.Y.), 
by the Japan-US Cooperative Research Program which is founded 
by the Japan Society for the Promotion of Science 
and the US National Science Foundation, 
and by the Grants-in Aid of the Ministry of Education, 
Science, Sports and Culture of Japan, 08640329 \& 10640228 (S.M.). 
\end{acknowledgements}

\end{document}